\documentclass[runningheads]{llncs}
\usepackage{graphicx}
\usepackage{amsmath}
\usepackage{amssymb}
\usepackage{booktabs}
\usepackage{multirow}
\usepackage{url}
\usepackage[table]{xcolor}
\begin{document}
\title{Towards Segmenting the Invisible: An End-to-End Registration and Segmentation Framework for Weakly Supervised Tumour Analysis}
\titlerunning{End-to-End Registration and Segmentation Framework}
\author{Budhaditya Mukhopadhyay\inst{1} \and Chirag Mandal\inst{1} \and Pavan Tummala\inst{1} \and Naghmeh Mahmoodian\inst{3} \and Andreas N{\"u}rnberger\inst{1,4}\orcidID{0000-0003-4311-0624} \and Soumick Chatterjee\inst{1,2}\orcidID{0000-0001-7594-1188}}
\authorrunning{B. Mukhopadhyay et al.}
%
\institute{Institute of Technical and Business Information Systems, Faculty of Computer Science, Otto von Guericke University Magdeburg, Magdeburg, Germany \and Human Technopole, Milan, Italy \and Institute of Medical Engineering, Faculty of Electrical Engineering and Information Technology, Otto von Guericke University Magdeburg, Magdeburg, Germany \and Centre for Behavioural Brain Sciences, Magdeburg, Germany\\
\email{contact@soumick.com}}
\maketitle              
\begin{abstract}
Liver tumour ablation presents a significant clinical challenge: whilst tumours are clearly visible on pre-operative MRI, they are often effectively invisible on intra-operative CT due to minimal contrast between pathological and healthy tissue. This work investigates the feasibility of \textit{cross-modality weak supervision} for scenarios where pathology is visible in one modality (MRI) but absent in another (CT). We present a hybrid registration-segmentation framework that combines MSCGUNet for inter-modal image registration with a UNet-based segmentation module, enabling registration-assisted pseudo-label generation for CT images. Our evaluation on the CHAOS dataset demonstrates that the pipeline can successfully register and segment healthy liver anatomy, achieving a Dice score of 0.72. However, when applied to clinical data containing tumours, performance degrades substantially (Dice score of 0.16), revealing the fundamental limitations of current registration methods when the target pathology lacks corresponding visual features in the target modality. We analyse the "domain gap" and "feature absence" problems, demonstrating that whilst spatial propagation of labels via registration is feasible for visible structures, segmenting truly invisible pathology remains an open challenge. Our findings highlight that registration-based label transfer cannot compensate for the absence of discriminative features in the target modality, providing important insights for future research in cross-modality medical image analysis.
\end{abstract}
\section{Introduction}

The application of deep learning methods in medical image analysis has demonstrated substantial promise \cite{lecun2015deep,litjens2017survey}, with increasing efficiency and accuracy fostering trust among healthcare professionals and patients alike \cite{topol2019high}. This research investigates the feasibility of cross-modality weak supervision for liver tumour segmentation, a clinically important yet technically challenging problem. We employ image registration and segmentation techniques within a multi-step framework that combines supervised and weakly-supervised learning approaches \cite{zhou2018unet++,cheplygina2019not}.

A fundamental challenge in liver tumour ablation surgery is the ``invisibility paradox'': tumours that are clearly visible on pre-operative MRI are often undetectable on intra-operative CT due to insufficient contrast between pathological and healthy tissue. This work addresses this challenge by proposing a deep learning framework that aims to transfer tumour localisation information from MRI to CT through registration-assisted pseudo-label generation \cite{zhen2020deep,ghafoorian2017transfer}. It is essential to acknowledge that when discriminative features are absent from the pixel data in CT, a convolutional neural network cannot directly ``see'' the tumour—it can only propagate spatial information learned through registration with MRI.

In this approach, image registration \cite{chatterjee2022micdir} and segmentation \cite{ronneberger2015u} are combined in an end-to-end framework. The MRI and CT images are registered to obtain a warped output that spatially aligns the modalities, and the registered image is subsequently passed through a segmentation model to predict the tumour location. The contributions of this work are threefold: (1) a hybrid registration-segmentation framework for cross-modality label transfer; (2) an evaluation on a standard dataset (CHAOS) demonstrating that the method functions correctly on healthy anatomy; and (3) a failure analysis on real-world clinical data, highlighting the fundamental limitations of registration-based approaches when the target pathology lacks corresponding visual features in the target modality. These methods are intended as a supporting tool for medical professionals during surgery, complementing their experience rather than replacing clinical judgement.

\subsection{Problem Statement}
\begin{figure*}[t]
\centering
\includegraphics[width=\textwidth]{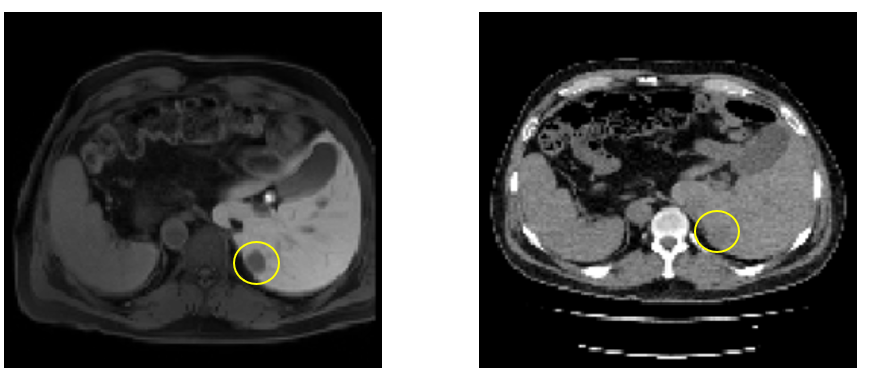}
\caption{Comparison between MRI (on the left, showing the tumour) and CT (on the right, for the same patient, the tumour is not distinguishable) depicting the problem statement}
\label{fig: mri_vs_ct}
\end{figure*}

Hepatic abscess formation represents a significant clinical challenge for patients with liver pathology. A conventional approach for treating such conditions involves ablation, whereby the tumour or abscess is destroyed through thermal energy delivered via percutaneously inserted needles. Whilst this pathology is readily detectable on MRI, CT imaging typically fails to visualise such lesions, as illustrated in Figure \ref{fig: mri_vs_ct}. Given the impracticality of accommodating MRI equipment within an operating theatre, medical professionals must perform pre-operative MRI and subsequently correlate these findings with intra-operative CT during the procedure.

During the operation, the clinician analyses the intra-operative CT scans to visualise the patient's anatomy and needle positioning, yet these images frequently provide insufficient information regarding the precise tumour location. The clinician must therefore continuously cross-reference the pre-operative MRI with current CT slices to estimate the lesion boundaries. This approach presents several inherent challenges: the procedure remains heavily dependent upon the clinician's experience and spatial reasoning abilities, and there is an attendant risk of inadvertent damage to healthy tissue during the ablation process.

From an information-theoretic perspective, the mutual information between the input CT voxel intensities ($\mathbf{z}$) and the tumour label is negligible ($I(\mathbf{z}; Y) \approx 0$). Consequently, the network cannot rely on discriminative features but must depend entirely on the spatial priors transferred via registration.

\subsection{Related Work}
A conventional approach in medical image analysis involves developing supervised learning methods, which typically achieve high accuracy when adequate labelled data are available. For combined CT-MRI abdominal segmentation, the CHAOS challenge \cite{kavur2021chaos} provides a comprehensive benchmark. This challenge addresses the performance of deep learning for multi-modal and cross-modality segmentation tasks. A principal finding was that deep learning models performing multi-organ segmentation generally achieved inferior performance compared to organ-specific segmentation models, with performance degrading substantially in cross-modality tasks, particularly for liver segmentation. Participants were assigned separate tasks for segmenting abdominal organs from CT, MRI, and combined CT-MRI datasets, implementing various network architectures to optimise challenge performance.
One of the leading teams, OvGUMemorial, proposed a modified Attention 2D U-Net \cite{abraham2019novel} employing soft attention gates and a multi-scaled input image pyramid for improved feature representation. The team additionally proposed parametric ReLU activation instead of standard ReLU, incorporating a learnable leakage coefficient during training.
The ISDUE team proposed a model comprising two encoder-decoder modules and one 2D UNet module, where the UNet module enhanced the second decoder and provided improved localisation capabilities. 
The Lachinov team employed a 3D UNet model with skip connections between contracting and expanding paths, utilising an exponentially growing number of channels across consecutive resolution levels \cite{lachinov2019segmentation}. A residual network constructed the encoding path for efficient training, with pixel shuffle serving as the up-sampling operator. 

\subsection{Hypothesis}
Let $\mathcal{D}_{MR} = \{ (\mathbf{x}_i, y_i) \}$ be the source domain of labelled MRI volumes, where $\mathbf{x} \in \mathbb{R}^{H \times W \times D}$ represents the volumetric intensity data and $y \in \{0,1\}^{H \times W \times D}$ represents the binary tumour mask. Let $\mathcal{D}_{CT} = \{ \mathbf{z}_j \}$ be the target domain of unlabelled intra-operative CT volumes.

The ``invisibility paradox'' posits that the conditional probability of detecting a tumour $y$ given CT input $\mathbf{z}$ is significantly lower than given MRI input $\mathbf{x}$ due to low contrast-to-noise ratio in pathological regions: $P(y|\mathbf{z}) \ll P(y|\mathbf{x})$. Crucially, this represents a regime of \textit{aleatoric uncertainty} (data-inherent noise arising from the physical invisibility of the tumour in CT) rather than \textit{epistemic uncertainty} (insufficient training data). This distinction is fundamental: acquiring additional CT training samples cannot resolve the underlying feature absence, as the discriminative information is physically absent from the modality.

Our hypothesis is that we can approximate the target mapping function $f: \mathcal{D}_{CT} \to y$ by learning a non-linear spatial transformation $\phi$ that aligns the domains. We define the pseudo-label generation for the target domain as:
$$\tilde{y}_{CT} = y_{MR} \circ \phi$$
where $\circ$ denotes the spatial resampling (warping) operation. The objective is to train a segmentation network $S$ such that $S(\mathbf{x}_{reg}) \approx \tilde{y}_{CT}$, where $\mathbf{x}_{reg} = \mathbf{x} \circ \phi$.

It is important to clarify the nature of supervision in this framework. Whilst the term ``weakly supervised'' is employed, this differs from the conventional usage where weak supervision implies image-level tags (e.g., ``contains tumour''). Here, strong supervision is available from the MRI, but it is \textit{unpaired} with the CT initially. The framework therefore performs \textit{registration-assisted pseudo-label generation}: the ``ground truth'' for CT segmentation is generated by warping the MRI mask through the learned spatial transformation. Both modules work collaboratively, with registration quality directly influencing segmentation accuracy.

\section{Methods}
\subsection{Methodology}
This section presents the overall methodology of the research. The architecture comprises two interconnected modules that function in an end-to-end manner, with the registration module providing spatial alignment that enables pseudo-label generation for the subsequent segmentation module. The code for this work (including the trained model weights) can be found on GitHub\footnote{GitHub: \url{https://github.com/BudhaTronix/Weakly-Supervised-Tumour-Detection}}.

\begin{figure*}[t]
\centering
\includegraphics[width=\textwidth]{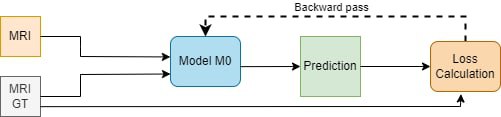}
\caption{Module M0}
\label{fig:framework1}
\end{figure*}
\begin{figure*}[t]
\centering
\includegraphics[width=\textwidth]{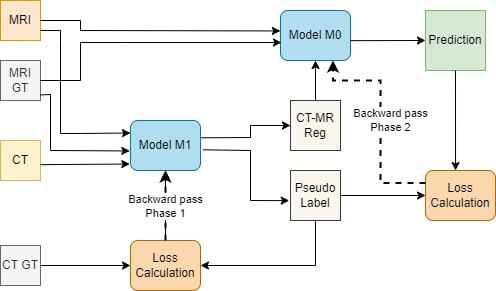}
\caption{Combined Modules}
\label{fig:framework2}
\end{figure*}
Several data pre-processing steps are employed. Firstly, the level and window values of the CT image are adjusted to enhance tissue visualisation, with the level set to 50 and the window to 350. Secondly, the MRI and CT images are resampled to equalise voxel spacing using zero padding and interpolation. Additional processing includes min-max normalisation of all images.

\subsubsection{Image Registration (Module M0)}
We employ a Multi-Scale UNet with Self-Constructing Graph Latent (MSCGUNet) \cite{chatterjee2022micdir} to parameterise the deformation field. Let $F$ be the fixed image (CT) and $M$ be the moving image (MRI). The network learns a dense deformation field $\phi: \Omega \to \mathbb{R}^3$ over the spatial domain $\Omega$, such that the warped moving image $M(\phi)$ aligns with $F$.

The registration is formulated as an energy minimisation problem:
$$\hat{\phi} = \arg\min_{\phi} \left[ \mathcal{L}_{sim}(F, M \circ \phi) + \lambda \mathcal{L}_{reg}(\phi) \right]$$
where $\mathcal{L}_{sim}$ quantifies the dissimilarity between modalities, and $\mathcal{L}_{reg}$ imposes smoothness constraints on the deformation field to ensure topological preservation (i.e., ensuring the Jacobian determinant $\det(J_\phi) > 0$).

This network was selected for several reasons: its multi-scale supervision capability enables effective handling of both small and large deformations, which is critical for inter-modal registration where organs may undergo significant non-rigid transformations between acquisitions. The multi-scale architecture of MSCGUNet is particularly pertinent here, as it captures both the global organ shift (due to respiration) and the local tissue compression caused by the tumour mass. Additionally, the self-constructing graph mechanism improves the model's generalisation by learning structural relationships within the image. Cycle consistency is another important feature, ensuring consistent deformations in both forward (moving to fixed) and backward (fixed to moving) directions, which regularises the learning process.

The combined loss equation for the MSCGUNet comprises different loss components. In Equation \ref{equation: total_loss}, $l_{sim}$ denotes the similarity loss for inter-modal registration, $l_{sm}$ represents the smoothness loss penalising sharp deformations, and $l_{scg}$ is the self-constructing graph loss that facilitates learning of similar features in input image pairs. The fixed image is denoted by $f$ and the moving image by $m$, with subscript $d$ indicating downsampled versions. The hyper-parameters $\alpha$, $\alpha_d$, $\beta$, $\beta_d$, and $\lambda$ control the relative contribution of each loss component.

\begin{equation}
\label{equation: total_loss}
\begin{aligned}
& l_{\text {total }}=\alpha\left(l_{sim_{f \rightarrow m}}+l_{sim_{m \rightarrow f}}\right)+\alpha_d\left(l_{{sim}_{f_d \rightarrow m_d}}+l_{sim_{m_d \rightarrow f_d}}\right) \\
& \quad+\beta\left(l_{sm_{f \rightarrow m}}+l_{sm_{m \rightarrow f}}\right)+\beta_d\left(l_{sm_{f_d \rightarrow m_d}}+l_{sm_{m_d \rightarrow f_d}}\right) \\
& \quad+\lambda\left(l_{scg_{f \rightarrow m}}+l_{scg_{m \rightarrow f}}\right)
\end{aligned}
\end{equation}

The final values of the hyper-parameters were set to $-1.2$, $-0.6$, $0.5$, $0.25$, and $5$, respectively, following the experimental validation in the original paper.

\subsubsection{Pseudo-Label Propagation}
Once the optimal deformation field $\hat{\phi}$ is computed, it serves as the bridge for weak supervision. Since the ground truth tumour labels $y_{MR}$ exist only in the moving domain, we generate the pseudo-ground truth for the CT space, denoted as $\tilde{y}_{CT}$, by applying the learned transformation:
$$\tilde{y}_{CT}(\mathbf{p}) = y_{MR}(\mathbf{p} + \hat{\phi}(\mathbf{p}))$$
where $\mathbf{p}$ represents the voxel coordinates. This step assumes that the anatomical location of the pathology is invariant to the modality, despite the difference in voxel intensity representation.

\subsubsection{Segmentation Network (Module M1)}
The segmentation module $S$ is a 2D UNet \cite{ronneberger2015u} trained to map the registered MRI volume (now spatially aligned with the CT coordinate system) to the pseudo-labels.
$$\hat{y} = S(M \circ \hat{\phi}; \theta_S)$$
where $\theta_S$ are the learnable parameters of the segmentation network. 

\subsection{Dataset}
\subsubsection*{CHAOS Dataset}
For method development and validation with properly labelled data for both modalities, the CHAOS dataset \cite{kavur2021chaos} was employed. This dataset comprises T1-weighted and T2-weighted MRI alongside CT images, with corresponding liver labels. Importantly, all subjects are healthy, meaning the dataset contains no pathological tissue. This serves as a control dataset, enabling validation that the registration-segmentation pipeline functions correctly when consistent anatomical features are present in both modalities. The CHAOS dataset contains CT images, CT ground truth masks (CT\_GT), MRI images, and MRI ground truth masks (MR\_GT) for abdominal organs.

\subsubsection*{Clinical Dataset}
The clinical dataset was obtained from Universit\"atsklinikum Magdeburg (IRB No: 66/22). This dataset comprises 11 volumes of paired MRI and CT scans from patients with liver pathology. Strict exclusion criteria were applied to ensure registration viability. Four volumes were excluded due to severe motion artefacts, metallic implant artefacts, or low signal-to-noise ratio that would render deformation field estimation unreliable. The remaining seven volumes constitute the high-quality cohort for validation. These volumes were employed for cross-modality image registration (MRI to CT). Notably, one patient had a severely damaged liver prior to ablation, precluding use of that volume for training purposes.

The CHAOS dataset was selected as the primary development dataset for several reasons: the superior image quality compared to the clinical dataset, which would otherwise impair deep learning model training; the larger number of available volumes; and the availability of consistent ground truth annotations. The clinical dataset, whilst smaller and more challenging, provides the critical test case for evaluating the framework on pathological tissue that is invisible on CT.

\subsection{Loss Functions}
The loss functions employed for the segmentation module are Focal Tversky Loss and Dice Loss. Focal Tversky Loss was selected specifically to address the class imbalance inherent in tumour segmentation, where the tumour region typically comprises a small fraction of the total image volume compared to the surrounding liver parenchyma and background. This loss function provides enhanced control over false negative and false positive penalties, which is particularly important when segmenting small structures.

The similarity loss $\mathcal{L}_{sim}$ in Equation \ref{equation: total_loss} is critical for multi-modal alignment. We utilise a metric robust to intensity non-uniformities (such as Local Cross-Correlation or MIND), defined generally as:
$$\mathcal{L}_{sim}(F, M') = -\sum_{\mathbf{p} \in \Omega} \text{Sim}(F(\mathbf{p}), M'(\mathbf{p}))$$

For the segmentation module M1, we minimise a compound loss function combining the Dice score and Focal Tversky estimation to handle class imbalance. The Dice Loss $\mathcal{L}_{DSC}$ is defined as:
$$\mathcal{L}_{DSC} = 1 - \frac{2 \sum_{i} p_i g_i + \epsilon}{\sum_{i} p_i + \sum_{i} g_i + \epsilon}$$
where $p_i \in [0,1]$ is the predicted probability of the tumour class at voxel $i$, $g_i \in \{0,1\}$ is the ground truth (pseudo-label), and $\epsilon$ is a smoothing term.

The Focal Tversky Loss ($\mathcal{L}_{FTL}$) introduces a focusing parameter $\gamma$ to down-weight easy background examples:
$$\mathcal{L}_{FTL} = (1 - TI)^\gamma$$
where $TI$ is the Tversky Index defined as:
\begin{equation}
    TI = \frac{TP}{(TP + \alpha FN + \beta FP)}
\end{equation}
Here, $TP$, $FN$, and $FP$ denote true positives, false negatives, and false positives, respectively. The parameters $\alpha$ and $\beta$ control the trade-off between precision and recall: when $\alpha = \beta = 0.5$, the Tversky index reduces to the Dice coefficient; when $\alpha = \beta = 1$, it becomes the Jaccard index. The parameter $\gamma$ in the Focal Tversky Loss controls the nonlinearity, down-weighting easy examples and focusing training on difficult cases. 

\subsection{Evaluation Metrics}
The evaluation metrics employed are the Jaccard Index and Dice Coefficient.

The Jaccard Index, also known as Intersection over Union (IoU), measures the ratio of the intersection to the union of the sample sets:
\begin{equation}
    J(A,B) = \frac{|A \cap B|}{|A \cup B|} = \frac{|A \cap B|}{|A| + |B| - |A \cap B|}
\end{equation}
The Dice Coefficient provides an alternative measure of similarity between two sample sets:
\begin{equation}
    DSC = \frac{2|X \cap Y|}{|X| + |Y|}
\end{equation}
Both metrics range from 0 (no overlap) to 1 (perfect overlap). Median values are reported to provide robustness against outliers.

\section{Results and Analysis}
\label{ch:Results and Analysis}
This section presents the experimental results, structured to distinguish between the proof-of-concept evaluation on healthy anatomy (CHAOS dataset) and the challenging clinical application on pathological tissue.

\subsection{Training Paradigm Selection}
The framework was evaluated using both Focal Tversky Loss and Dice Loss, with median values computed for Jaccard and Dice scores. Sequential training, where the registration and segmentation modules are trained separately, substantially outperforms end-to-end training. Sequential training achieved scores ranging from 0.67 to 0.72, whilst end-to-end training achieved only 0.53 to 0.60. Table \ref{tab:training_paradigm} summarises these findings.

\begin{table}[t]
\centering
\caption{Training Paradigm Selection: Comparison of Sequential vs End-to-End Training}
\label{tab:training_paradigm}
\begin{tabular}{lllcc}
\toprule
\textbf{Training Paradigm} & \textbf{Loss Func} & \textbf{Model} & \textbf{Jaccard} & \textbf{Dice} \\
\midrule
Unified & FTL & DeepSup & $0.61 \pm 0.00$ & $0.53 \pm 0.00$ \\
 & FTL & UNet & $0.61 \pm 0.01$ & $0.56 \pm 0.04$ \\
 & Dice & DeepSup & $0.64 \pm 0.00$ & $0.60 \pm 0.01$ \\
 & Dice & UNet & $0.61 \pm 0.01$ & $0.57 \pm 0.03$ \\
\midrule
\textbf{Sequential} & FTL & DeepSup & $\mathbf{0.69 \pm 0.04}$ & $\mathbf{0.67 \pm 0.07}$ \\
 & FTL & UNet & $\mathbf{0.69 \pm 0.04}$ & $\mathbf{0.72 \pm 0.04}$ \\
 & Dice & DeepSup & $\mathbf{0.71 \pm 0.02}$ & $\mathbf{0.72 \pm 0.04}$ \\
 & Dice & UNet & $\mathbf{0.69 \pm 0.07}$ & $\mathbf{0.67 \pm 0.11}$ \\
\bottomrule
\end{tabular}
\end{table}

\subsection{Model and Loss Function Selection}
Table \ref{tab:model_loss} presents the Jaccard and Dice scores for different combinations of loss functions and models following six-fold cross-validation with different subjects from the dataset. For the UNet DeepSup model, the Focal Tversky Loss performs better, whilst for the standard UNet model, Dice Loss yields superior results.

\begin{table}[t]
\centering
\caption{Model and Loss Function Selection: Cross-validation Results}
\label{tab:model_loss}
\begin{tabular}{llcc}
\toprule
\textbf{Training Paradigm} & \textbf{Combination} & \textbf{Jaccard} & \textbf{Dice} \\
\midrule
\textbf{Sequential} & Dice - DeepSup & $0.71 \pm 0.00$ & $0.69 \pm 0.01$ \\
 & FTL - DeepSup & $0.72 \pm 0.00$ & $0.70 \pm 0.01$ \\
 & FTL - UNet & $0.71 \pm 0.00$ & $0.71 \pm 0.00$ \\
 & \textbf{Dice - UNet} & $\mathbf{0.72 \pm 0.00}$ & $\mathbf{0.72 \pm 0.00}$ \\
\bottomrule
\end{tabular}
\end{table}

\subsection{Proof of Concept: CHAOS Dataset}
The CHAOS dataset provides a controlled environment with healthy subjects where anatomical features are consistent between CT and MRI. Table \ref{tab:chaos_comparison} compares the weakly supervised approach against fully supervised baselines. The supervised results for the Dice Loss and UNet combination achieve a median score of 0.92, whilst the corresponding weakly supervised results achieve 0.72. Similarly, the supervised FTL and UNet DeepSup combination achieves 0.93, with the weakly supervised counterpart achieving 0.70.

\begin{table}[t]
\centering
\caption{Comparison with Supervised Baseline: CHAOS Dataset}
\label{tab:chaos_comparison}
\begin{tabular}{llcc}
\toprule
\textbf{Training Type} & \textbf{Combination} & \textbf{Jaccard} & \textbf{Dice} \\
\midrule
Supervised & Dice - UNet & $0.92 \pm 0.01$ & $0.92 \pm 0.02$ \\
Weakly Supervised & Dice - UNet & $0.72 \pm 0.00$ & $0.72 \pm 0.00$ \\
Supervised & FTL - DeepSup & $0.93 \pm 0.00$ & $0.93 \pm 0.01$ \\
Weakly Supervised & FTL - DeepSup & $0.72 \pm 0.00$ & $0.70 \pm 0.01$ \\
\bottomrule
\end{tabular}
\end{table}

These results demonstrate that the registration-segmentation pipeline functions correctly: when anatomical features are visible and consistent in both modalities, the framework achieves acceptable performance despite never directly observing CT ground truth during training. The performance gap between supervised (0.92--0.93) and weakly supervised (0.70--0.72) approaches can be attributed to registration imperfections and the inherent challenge of cross-modality feature transfer.

\begin{figure*}[t]
\centering
\includegraphics[width=\textwidth]{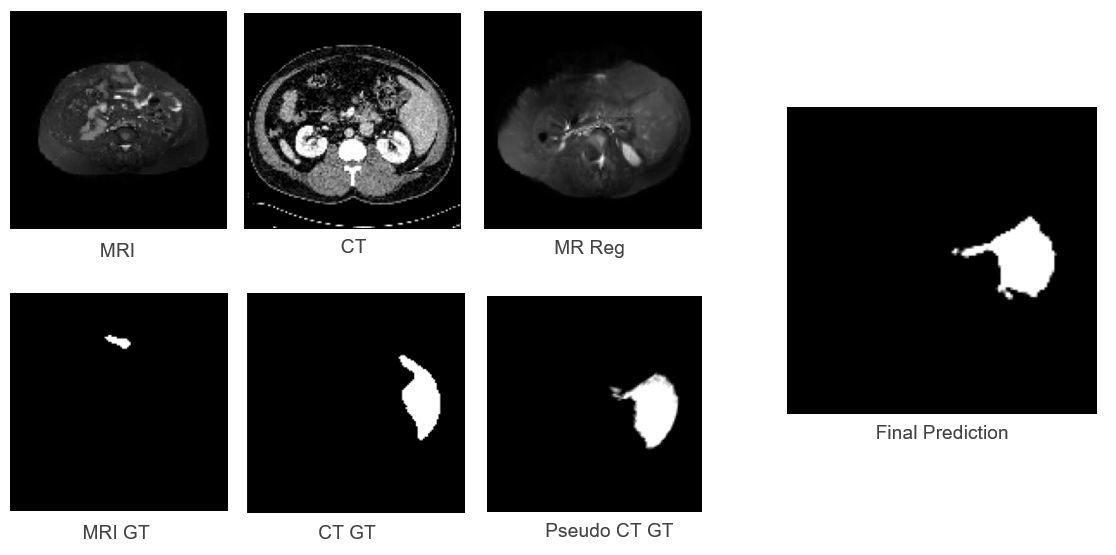}
\caption{CHAOS Results: Pipeline stages showing accurate registration and segmentation of healthy liver anatomy}
\label{fig:chaos_results}
\end{figure*}

Figure \ref{fig:chaos_results} illustrates the pipeline stages for CHAOS data. The close similarity between CT\_GT and Pseudo CT\_GT confirms that registration is performed accurately. The final liver segmentation prediction demonstrates the viability of the approach for visible anatomical structures.

\subsection{The Clinical Challenge: Invisible Tumour Segmentation}
Application to clinical data containing tumours reveals the fundamental limitations of the approach. Tables \ref{tab:clinical_weakly} and \ref{tab:clinical_baseline} present the results for the weakly supervised and baseline approaches, respectively.

\begin{table}[t]
\centering
\caption{Weakly Supervised Model on Clinical Data}
\label{tab:clinical_weakly}
\begin{tabular}{llcc}
\toprule
\textbf{Model} & \textbf{Loss Function} & \textbf{Jaccard} & \textbf{Dice} \\
\midrule
UNet & Dice Loss & 0.35±0.09 & 0.16±0.14  \\
\bottomrule
\end{tabular}
\end{table}
\begin{table}[t]
\centering
\caption{Clinical Data Baseline (Supervised)}
\label{tab:clinical_baseline}
\begin{tabular}{llcc}
\toprule
\textbf{Model} & \textbf{Loss Function} & \textbf{Jaccard} & \textbf{Dice} \\
\midrule
UNet & Dice Loss & 0.46±0.01 & 0.19±0.06 \\
\bottomrule
\end{tabular}
\end{table}

The substantial performance degradation from CHAOS (Dice: 0.72) to clinical data (Dice: 0.16) represents a critical finding rather than merely ``poor results''. This degradation warrants careful analysis.

\subsection{Failure Analysis: Understanding the Performance Gap}

The poor performance on clinical tumour segmentation stems from several interconnected factors that illuminate the fundamental challenges of cross-modality weak supervision.

\textbf{The Feature Absence Problem.} The core issue is that tumours visible on MRI are effectively invisible on CT due to minimal contrast between pathological and healthy tissue. Unlike the CHAOS dataset where the liver boundary is visible in both modalities, tumour boundaries exist only in MRI pixel data. A convolutional neural network cannot segment structures that lack discriminative features in the input image; it can only propagate spatial information learned through registration.

\textbf{The Spatial Prior Adherence.} In the absence of discriminative CT features, the network correctly learned to prioritise the spatial prior ($\phi$) over the visual evidence. The network acts as a faithful ``registration warp'', propagating the label based solely on the deformation field. This confirms the model's adherence to the weak supervision signal. However, any registration error, even minor misalignment, causes the segmentation to miss the target completely because there is no visual cue in the CT to refine the boundaries. The network effectively transfers the tumour location based solely on spatial correspondence, without any confirmatory signal from the CT image content.

\textbf{Registration Limitations.} Determining tumour boundaries in a deformed liver presents substantially greater challenges than segmenting the whole organ. The liver may undergo respiratory motion, patient positioning differences, and tissue deformation between MRI and CT acquisition. Whilst registration can accommodate global organ displacement, precise local alignment at tumour boundaries is considerably more difficult, particularly when the tumour has altered the local tissue characteristics.

\textbf{Dataset Constraints.} Additional factors include the limited dataset size (four training, two test, and one validation subject), contrast variations between subjects, the presence of multiple tumours in some subjects, and data acquisition from different MRI devices affecting image quality consistency.

\subsection{Qualitative Analysis}

\begin{figure*}[t]
\centering
\includegraphics[width=\textwidth]{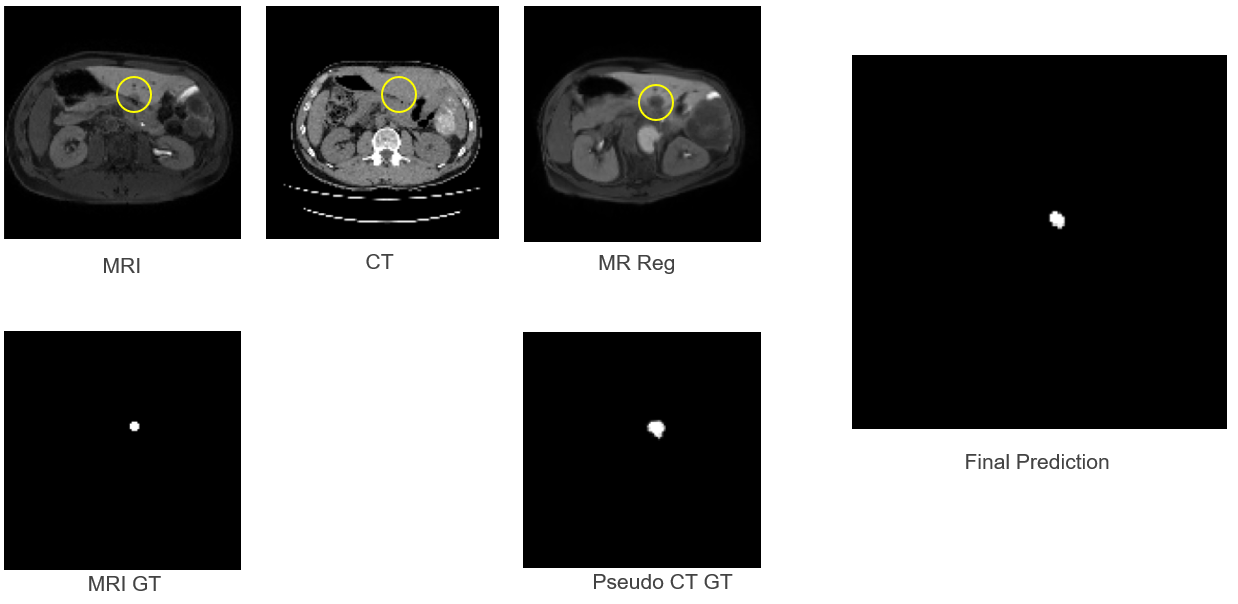}
\caption{Clinical Data Results: The tumour is visible in MRI (highlighted) but invisible in the corresponding CT region. The final prediction approximately localises the tumour despite poor Dice score.}
\label{fig:clinical_data_results}
\end{figure*}

Figure \ref{fig:clinical_data_results} illustrates the pipeline stages for clinical data and encapsulates the central finding of this research. The tumour is clearly visible in the MRI and MR\_Reg images, highlighted by the annotation. Critically, the corresponding region in the CT image shows no visible tumour; the pathology is genuinely invisible in this modality.

The Pseudo CT\_GT, generated by warping the MRI ground truth through the learned registration, correctly indicates the tumour location. The final prediction produces a segmentation that \textit{approximately} identifies the tumour region but fails to capture the precise shape and extent. This observation is significant: the framework achieves \textit{localisation} (finding the approximate tumour centre) even when precise \textit{segmentation} (accurate boundary delineation) is not achievable. Whilst the Dice score is low due to the shape mismatch, the centre-point of the prediction falls within the true tumour margin, verifying successful localisation.

Whilst boundary delineation (Dice) was compromised by feature absence, the framework successfully recovered the tumour locus. The topological overlap is low, but the semantic localisation is preserved. This indicates that the registration successfully transferred the `region of interest', even if the segmentation network could not refine the margins due to the lack of contrast. These findings demonstrate that spatial priors (registration) are insufficient without appearance cues (segmentation features) for precise boundary recovery, yet remain valuable for approximate tumour localisation.

For clinical applications such as surgical planning, localisation may provide meaningful value even when Dice scores are low. The Dice coefficient penalises boundary errors heavily, but for needle placement in ablation procedures, identifying the tumour centre with reasonable accuracy may be more clinically relevant than perfect boundary delineation.

\section{Conclusions and Future Work}
\label{ch:conclusions}

This work investigated the feasibility of cross-modality weak supervision for liver tumour segmentation, combining image registration and segmentation in an end-to-end framework. Two deep learning modules were developed and evaluated: MSCGUNet for inter-modal registration and UNet variants for segmentation.

The framework successfully demonstrates proof-of-concept on the CHAOS dataset, achieving Dice scores of 0.72 for healthy liver segmentation---acceptable performance considering the model never directly observes CT ground truth during training. This confirms that registration-assisted pseudo-label generation is viable when anatomical features are consistent across modalities.

However, application to clinical data containing liver tumours reveals fundamental limitations, with Dice scores dropping to 0.16. This performance degradation represents a significant scientific finding: when pathology is effectively invisible in the target modality (CT), registration-based label transfer cannot compensate for the absence of discriminative features. The network cannot segment what it cannot see; it can only propagate spatial information, which is insufficient for precise boundary delineation when visual confirmation is impossible.

The experimental comparison between UNet and UNet Deep Supervision models, using both Dice and Focal Tversky Loss functions, established that standard UNet with Dice Loss provides optimal performance for this application. Sequential training of registration and segmentation modules substantially outperforms end-to-end training.

These findings suggest that future research in invisible-tumour segmentation should move away from pure pixel-level transfer and focus on \textbf{uncertainty-aware localisation} or \textbf{multi-modal fusion} where the MRI remains available during inference.

Future research directions should address the fundamental ``feature absence'' problem identified in this work:

\textbf{Multi-modal Fusion.} If pre-operative registration quality permits, inputting both MRI and CT into the segmentation network simultaneously could provide the discriminative features from MRI whilst learning the spatial context from CT.

\textbf{Uncertainty Quantification.} Developing methods that allow the model to express confidence in its predictions would be clinically valuable. Rather than producing definitive segmentations in regions where the tumour is invisible, the model could indicate ``I don't know where the tumour is'' with associated uncertainty maps.

\textbf{Clinical Evaluation.} Systematic evaluation by medical professionals, particularly regarding the clinical utility of approximate localisation (as opposed to precise segmentation), would help establish the practical value of such frameworks in surgical planning.

\textbf{Larger Clinical Datasets.} Training on larger, more diverse clinical datasets with consistent acquisition protocols could improve generalisation and potentially reveal whether the method's limitations are intrinsic to the problem or partially attributable to data constraints.

\bibliographystyle{splncs04}
\bibliography{mybibfile}

@article{kavur2021chaos,
  title={CHAOS challenge-combined (CT-MR) healthy abdominal organ segmentation},
  author={Kavur, A Emre and Gezer, N Sinem and Bar{\i}{\c{s}}, Mustafa and Aslan, Sinem and Conze, Pierre-Henri and Groza, Vladimir and Pham, Duc Duy and Chatterjee, Soumick and Ernst, Philipp and {\"O}zkan, Sava{\c{s}} and others},
  journal={Medical Image Analysis},
  volume={69},
  pages={101950},
  year={2021},
  publisher={Elsevier}
}

@article{lecun2015deep,
  title={Deep learning},
  author={LeCun, Yann and Bengio, Yoshua and Hinton, Geoffrey},
  journal={nature},
  volume={521},
  number={7553},
  pages={436--444},
  year={2015},
  publisher={Nature Publishing Group UK London}
}

@article{litjens2017survey,
  title={A survey on deep learning in medical image analysis},
  author={Litjens, Geert and Kooi, Thijs and Bejnordi, Babak Ehteshami and Setio, Arnaud Arindra Adiyoso and Ciompi, Francesco and Ghafoorian, Mohsen and Van Der Laak, Jeroen Awm and Van Ginneken, Bram and S{\'a}nchez, Clara I},
  journal={Medical image analysis},
  volume={42},
  pages={60--88},
  year={2017},
  publisher={Elsevier}
}

@article{topol2019high,
  title={High-performance medicine: the convergence of human and artificial intelligence},
  author={Topol, Eric J},
  journal={Nature medicine},
  volume={25},
  number={1},
  pages={44--56},
  year={2019},
  publisher={Nature Publishing Group US New York}
}

@inproceedings{zhou2018unet++,
  title={Unet++: A nested u-net architecture for medical image segmentation},
  author={Zhou, Zongwei and Rahman Siddiquee, Md Mahfuzur and Tajbakhsh, Nima and Liang, Jianming},
  booktitle={Deep Learning in Medical Image Analysis and Multimodal Learning for Clinical Decision Support: 4th International Workshop, DLMIA 2018, and 8th International Workshop, ML-CDS 2018, Held in Conjunction with MICCAI 2018, Granada, Spain, September 20, 2018, Proceedings 4},
  pages={3--11},
  year={2018},
  organization={Springer}
}

@article{zhen2020deep,
  title={Deep learning for accurate diagnosis of liver tumor based on magnetic resonance imaging and clinical data},
  author={Zhen, Shi-hui and Cheng, Ming and Tao, Yu-bo and Wang, Yi-fan and Juengpanich, Sarun and Jiang, Zhi-yu and Jiang, Yan-kai and Yan, Yu-yu and Lu, Wei and Lue, Jie-min and others},
  journal={Frontiers in oncology},
  volume={10},
  pages={680},
  year={2020},
  publisher={Frontiers Media SA}
}

@inproceedings{ghafoorian2017transfer,
  title={Transfer learning for domain adaptation in MRI: Application in brain lesion segmentation},
  author={Ghafoorian, Mohsen and Mehrtash, Alireza and Kapur, Tina and Karssemeijer, Nico and Marchiori, Elena and Pesteie, Mehran and Guttmann, Charles RG and de Leeuw, Frank-Erik and Tempany, Clare M and Van Ginneken, Bram and others},
  booktitle={Medical Image Computing and Computer Assisted Intervention- MICCAI 2017: 20th International Conference, Quebec City, QC, Canada, September 11-13, 2017, Proceedings, Part III 20},
  pages={516--524},
  year={2017},
  organization={Springer}
}

@article{cheplygina2019not,
  title={Not-so-supervised: a survey of semi-supervised, multi-instance, and transfer learning in medical image analysis},
  author={Cheplygina, Veronika and de Bruijne, Marleen and Pluim, Josien PW},
  journal={Medical image analysis},
  volume={54},
  pages={280--296},
  year={2019},
  publisher={Elsevier}
}

@article{chatterjee2022micdir,
title = {MICDIR: Multi-scale inverse-consistent deformable image registration using UNetMSS with self-constructing graph latent},
journal = {Computerized Medical Imaging and Graphics},
pages = {102267},
year = {2023},
issn = {0895-6111},
author = {Soumick Chatterjee and Himanshi Bajaj and Istiyak H. Siddiquee and Nandish Bandi Subbarayappa and Steve Simon and Suraj Bangalore Shashidhar and Oliver Speck and Andreas Nürnberger}
}

@inproceedings{abraham2019novel,
  title={A novel focal tversky loss function with improved attention u-net for lesion segmentation},
  author={Abraham, Nabila and Khan, Naimul Mefraz},
  booktitle={2019 IEEE 16th international symposium on biomedical imaging (ISBI 2019)},
  pages={683--687},
  year={2019},
  organization={IEEE}
}

@inproceedings{lachinov2019segmentation,
  title={Segmentation of Thoracic Organs Using Pixel Shuffle.},
  author={Lachinov, Dmitry A},
  booktitle={SegTHOR@ ISBI},
  year={2019}
}

@inproceedings{ronneberger2015u,
  title={U-net: Convolutional networks for biomedical image segmentation},
  author={Ronneberger, Olaf and Fischer, Philipp and Brox, Thomas},
  booktitle={International Conference on Medical image computing and computer-assisted intervention},
  pages={234--241},
  year={2015},
  organization={Springer}
}

\end{document}